\documentclass[11pt,a4paper]{article}
\usepackage{amsmath,amsfonts,amssymb,amsthm,mathtools}
\usepackage{authblk}
\usepackage{latexsym,graphicx}
\usepackage{dsfont}
\usepackage{amscd}
\usepackage[dvipsnames]{xcolor}
\usepackage{extarrows} 
\numberwithin{equation}{section} 
\usepackage{yfonts}
\usepackage{authblk}
\usepackage{appendix}
\usepackage{comment}
\usepackage[colorlinks=true]{hyperref}
\hypersetup{urlcolor=blue, citecolor=red, linkcolor=blue}

\usepackage[square,numbers]{natbib}
\allowdisplaybreaks

\newtheorem{theorem}{Theorem}

\newtheorem{proposition}{Proposition}[section]
\theoremstyle{remark}

\newcommand{\ii}{\mathrm{i}}

\newcommand{\C}{{\mathbb C}}
\newcommand{\Z}{{\mathbb Z}}

\newcommand{\tr}{\mathrm{tr}}

\newcommand{\cN}{\mathcal {N}}

\newcommand{\cV}{\mathcal{V}}

\newcommand{\mA}{\mathsf{A}}

\newcommand{\mC}{\mathsf{C}}

\newcommand{\mP}{\mathsf{P}}

\newcommand{\mR}{\mathsf{R}}

\title{Consistency of the B\"{a}cklund transformation for the spin Calogero-Moser system}
\date{}
\author{Bjorn K. Berntson}

\affil{Department of Physics, KTH Royal Institute of Technology, Stockholm, Sweden \\
Email: bbernts@kth.se}

\begin{document}

\maketitle

\begin{abstract}
We prove the consistency of the B\"{a}cklund transformation (BT) for the spin Calogero-Moser (sCM) system in the rational, trigonometric, and hyperbolic cases.  The BT for the sCM system consists of an overdetermined system of ordinary differential equations; to establish our result, we construct and analyze certain functions that measure the departure of this overdetermined system from consistency and show, under mild assumptions, that these functions are identically zero and that this allows for a unique solution to the initial value problem for the overdetermined system. 
\end{abstract} 



\section{Introduction}

Spin Calogero-Moser (sCM) systems describe arbitrary numbers of particles carrying internal degrees of freedom and interacting in one dimension \cite{gibbons1984,wojciechowski1985,krichever1995}. These systems preserve several integrability properties of the spinless Calogero-Moser systems they generalize; in the classical setting, one such property is the existence of a B\"{a}cklund transformation (BT) relating certain distinct solutions of sCM systems \cite{gibbons1983}. The BT for the sCM system has recently appeared in the form of discrete-time evolution equations for the sCM system \cite{zabrodin2019,prokofev2020} and has been employed in the construction of soliton solutions for spin generalizations of the Benjamin-Ono equation \cite{berntsonlangmann2022}.  

As will be delineated below, the BT for the sCM system consists of an overdetermined system of ordinary differential equations (ODEs), whose consistency was not elaborated in \cite{gibbons1983}. This note provides a direct proof that the BT for the sCM system is consistent under mild assumptions.

We consider complex sCM systems in the rational (case I), trigonometric (case II), and hyperbolic (case III) cases; corresponding results for the elliptic case \cite{krichever1995} are more complicated and will be presented elsewhere \cite{berntsonlangmann2022elliptic}. The cases we consider are distinguished by the following special functions which will appear as two-body interaction potentials in the sCM equations of motion below,
\begin{equation}\label{eq:V}
V(z)\coloneqq \begin{cases}
1/z^2  & \text{(case I)} \\
(\pi/L)^2/\sin^2(\pi z/L)&  \text{(case II)} \\
(\pi/2\delta)^2/\sinh^2(\pi z/2\delta) & \text{(case III)},
\end{cases}
\end{equation}
with $L>0$ and $\delta>0$ arbitrary parameters.  Let $N\in \Z_{\geq 1}$; the $j$th particle in an $N$-body sCM system is characterized by a time-dependent position $a_j(t)\in\C$ and internal degrees of freedom $|e_j(t)\rangle\in \cV$ and $\langle f_j(t)|\in \cV^*$ where $\cV$ is a $d$-dimensional complex vector space with $d\in \Z_{\geq 1}$ and $\cV^*$ is the corresponding dual vector space. To write the sCM equations of motion, we use Dirac bra-ket notation \cite{dirac1939} where $\langle f|e\rangle \in \C$ and $|e\rangle\langle f|\in \cV\otimes \cV^*\cong \C^{d\times d}$ are the inner and outer products, respectively. Then, in each case I--III, the sCM system is defined by the following system of equations,
\begin{equation}\label{eq:sCM1a}
\ddot{a}_j=-4\sum_{k\neq j}^N \langle f_j|e_k\rangle\langle f_k|e_j\rangle V'(a_j-a_k) \quad (j=1,\ldots,N)
\end{equation}
and
\begin{equation}\label{eq:sCM1b}
\begin{split}
|\dot{e}_j\rangle =&\; 2\ii\sum_{k\neq j}^N |e_k\rangle\langle f_k|e_j\rangle V(a_j-a_k), \\ 
\langle \dot{f}_j|=&\; -2\ii \sum_{k\neq j}^N \langle f_j|e_k\rangle\langle f_k|V(a_j-a_k)
\end{split} \quad (j=1,\ldots,N)
\end{equation}
and
\begin{equation}\label{eq:ejfj}
\langle e_j|f_j\rangle =1 \quad (j=1,\ldots,N),
\end{equation}
where dots above a variable indicate differentiation with respect to time, $V(z)$ is as in \eqref{eq:V}, and $\sum_{k\neq j}^N\coloneqq \sum_{k=1,k\neq j}^N$. We emphasize that \eqref{eq:ejfj} is only a constraint on the initial conditions for $\{|e_j\rangle,\langle f_j|\}_{j=1}^N$; if \eqref{eq:ejfj} is satisfied at $t=0$, a short calculation using \eqref{eq:sCM1b} shows that it holds at future times. 

Under certain conditions, a BT for the sCM system \cite{gibbons1983} links solutions of \eqref{eq:sCM1a}--\eqref{eq:ejfj} to those of a second sCM system with $M\in \Z_{\geq 1}$ particles,
\begin{equation}\label{eq:sCM2a}
\ddot{b}_j=-4\sum_{k\neq j}^M \langle h_j|g_k\rangle\langle h_k|g_j\rangle V'(b_j-b_k) \quad (j=1,\ldots,M)
\end{equation}
and
\begin{equation}\label{eq:sCM2b}
\begin{split}
|\dot{g}_j\rangle =&\; 2\ii\sum_{k\neq j}^N |g_k\rangle\langle h_k|g_j\rangle V(b_j-b_k), \\ 
\langle \dot{h}_j|=&\; -2\ii \sum_{k\neq j}^N \langle h_j|g_k\rangle\langle h_k|V(b_j-b_k)
\end{split} \quad (j=1,\ldots,M)
\end{equation}
and
\begin{equation}\label{eq:gjhj}
\langle h_j|g_j\rangle =1 \quad (j=1,\ldots,M).
\end{equation}
The BT between \eqref{eq:sCM1a}--\eqref{eq:ejfj} and \eqref{eq:sCM2a}--\eqref{eq:gjhj} reads
\begin{equation}\label{eq:BT}
\begin{split}
\dot{a}_j \langle f_j| =&\; 2\ii\sum_{k\neq j}^N \langle f_j|e_k\rangle\langle f_k|    \alpha(a_j-a_k)-2\ii\sum_{k=1}^M \langle f_j|g_k\rangle\langle h_k|\alpha(a_j-b_k) \quad (j=1,\ldots,N),    \\
\dot{b}_j |g_j\rangle =&\; -2\ii \sum_{k\neq j}^M |g_k\rangle\langle h_k|g_j\rangle \alpha(b_j-b_k)+2\ii\sum_{k=1}^N |e_k\rangle\langle f_k|g_j\rangle \alpha(b_j-a_k) \quad (j=1,\ldots,M),
\end{split}
\end{equation}
where
\begin{equation}\label{eq:alpha}
\alpha(z)\coloneqq \begin{cases}
1/z  & \text{(case I)} \\
(\pi/L)\cot(\pi z/L)    &  \text{(case II)} \\
(\pi/2\delta)\coth(\pi z/2\delta) & \text{(case III)},
\end{cases}
\end{equation}
and its significance is captured in the following proposition \cite{gibbons1983,berntsonlangmann2022}.

\begin{proposition}[BT for the sCM system]\label{prop:BT}
In each case I--III, the first-order equations \eqref{eq:sCM1b}, \eqref{eq:sCM2b}, and \eqref{eq:BT}, together with the constraints \eqref{eq:ejfj} and \eqref{eq:gjhj}, imply the second-order equations \eqref{eq:sCM1a} and \eqref{eq:sCM2a}.
\end{proposition}

The first-order system of equations \eqref{eq:sCM1b}, \eqref{eq:sCM2b}, and \eqref{eq:BT} in Proposition~\ref{prop:BT} is overdetermined, owing to the fact that there are $d$ independent equations for the time evolution of each of $a_j$ and $b_j$ in \eqref{eq:BT}, and thus is not obviously consistent. However, by right-multiplying the first set of equations in \eqref{eq:BT} by $|e_j\rangle$ and left-multiplying the second set of equations in \eqref{eq:BT} by $\langle h_j|$ and using \eqref{eq:ejfj} and \eqref{eq:gjhj}, we obtain
\begin{equation}\label{eq:BTscalar}
\begin{split}
\dot{a}_j  =&\; 2\ii\sum_{k\neq j}^N \langle f_j|e_k\rangle\langle f_k|e_j\rangle    \alpha(a_j-a_k)-2\ii\sum_{k=1}^M \langle f_j|g_k\rangle\langle h_k|e_j\rangle \alpha(a_j-b_k) \quad (j=1,\ldots,N),    \\
\dot{b}_j =&\; -2\ii \sum_{k\neq j}^M \langle h_j|g_k\rangle\langle h_k|g_j\rangle \alpha(b_j-b_k)+2\ii\sum_{k=1}^N \langle h_j |e_k\rangle\langle f_k|g_j\rangle \alpha(b_j-a_k) \quad (j=1,\ldots,M),
\end{split}
\end{equation}
which together with \eqref{eq:sCM1b} and \eqref{eq:sCM2b} forms a manifestly consistent system. We will use this consistent system to establish our main result: that under mild assumptions, the first-order equations of Proposition~\ref{prop:BT} admit a unique solution.

\begin{theorem}[Consistency of the BT for the sCM system]\label{thm:main}
In each case I--III, the initial value problem consisting of the first-order equations \eqref{eq:sCM1b}, \eqref{eq:sCM2b}, and \eqref{eq:BT} with initial conditions satisfying \eqref{eq:ejfj}, \eqref{eq:gjhj}, and \eqref{eq:BT} at $t=0$ has a unique solution on a maximal interval $[0,\tau)$, for some $\tau\in (0,\infty)$, where each solution variable remains finite in finite time and
\begin{equation}\label{eq:ajak}
\begin{gathered}
a_j-a_k \neq 0 \quad (1\leq j<k\leq N),\qquad b_j-b_k \neq 0 \quad (1\leq j<k\leq M), \\
a_j-b_k\neq 0 \quad (j=1,\ldots,N; k=1,\ldots,M)
\end{gathered}
\end{equation}
(with the equalities modulo $L$ and $2\ii\delta$ in cases II and III, respectively) holds.
\end{theorem}

We now give some remarks on Theorem~\ref{thm:main} and its proof, which is given below in Section~\ref{sec:proofs}. Certain basic properties of the special functions $\alpha(z)$ and $V(z)$ are recalled in Appendix~\ref{app:special}. 

\subsection{Remarks on the result}

\begin{enumerate}

\item We prove cases I--III concurrently. Only properties of the special functions $\alpha(z)$ and $V(z)$ that are the same in all three cases are needed in the proof.

\item Given initial data satisfying \eqref{eq:ejfj}, \eqref{eq:gjhj}, and \eqref{eq:BT}, we may solve the consistent system \eqref{eq:sCM1b}, \eqref{eq:sCM2b}, and \eqref{eq:BTscalar} to determine $\{a_j,|e_j\rangle,\langle f_j|\}_{j=1}^{N}$ and $\{b_j,|g_j\rangle,\langle h_j|\}_{j=1}^{M}$ on a maximal interval $[0,\tau)$. The key idea in our proof of Theorem~\ref{thm:main} is to show that, on such a solution, a set of $\cV\otimes \cV^*$-valued functions $\{\mR_j\}_{j=1}^{N+M}$ (see \eqref{eq:Rj} for the definition, which uses notation introduced in Section~\ref{subsec:notation}) that measure the differences between the left- and right-hand sides of \eqref{eq:BT} satisfy a system of linear homogeneous ODEs (see \eqref{eq:Rjdot2}). Because $\mR_j(0)=0$ for $j=1,\ldots,N+M$, as the initial data is assumed to satisfy \eqref{eq:BT}, it follows that the unique solution of the linear homogeneous system is $\{\mR_j(t)=0\}_{j=1}^{N+M}$ on $[0,\tau)$, which implies that \eqref{eq:BT} holds on $[0,\tau)$. 

\item We use the concept of a maximal solution of a system of ODEs (see, e.g., \cite[Corollary~3.2]{hartman1982})   in Theorem~\ref{thm:main} and its proof. Given an initial value problem for a system of ODEs $\dot{y}_j=F_j(y_1,\ldots,y_N)$, $j=1,\ldots,N$, where each $F_j$ is Lipschitz near the initial data imposed at $t=0$, a local solution exists by the Picard-Lindel\"{o}f theorem. This solution may be extended as long as (i) each variable $y_j$ remains finite in finite time and (ii) each function $F_j$ remains Lipschitz near the solution, up to a maximal time $\tau\in (0,\infty)$. The resulting maximal solution is unique on the maximal interval $[0,\tau)$ (see, e.g., \cite[Theorem~8.1]{hartman1982}). For the system we consider, the condition \eqref{eq:ajak} guarantees (ii) is satisfied.

\item The BT for the sCM system was proposed in \cite{gibbons1983} in the case $N=M$. Our conventions for the B\"{a}cklund transformation \eqref{eq:BT} (with $N=M$) are different from but equivalent to those in \cite[Eq.~(17)]{gibbons1983} via the transformations $t\to -2t$ and
\begin{equation}
(a_j,\dot{a}_j,|e_j\rangle,\langle f_j|,b_j,\dot{b}_j,|g_j\rangle,\langle h_j|) \to (x_j^+,p_j^+,|e_j^+ \rangle,\langle f_j^+|,x_j,p_j,|e_j\rangle,\langle f_j|) \quad (j=1,\ldots,N).
\end{equation}

\end{enumerate}

\section{Proof of Theorem~\ref{thm:main}}\label{sec:proofs}

We first introduce shorthand notation used in the proof in Section~\ref{subsec:notation}. Theorem~\ref{thm:main} is proved in Section~\ref{sec:mainproof}.

\subsection{Notation}\label{subsec:notation}
We make the following definitions,
\begin{equation}\label{eq:SH}
(a_j,|e_j\rangle,\langle f_j|,r_j)\coloneqq \begin{cases}
(a_j,|e_j\rangle,\langle f_j| ,+1) & j=1,\ldots,N \\
(b_j,|g_j\rangle,\langle h_j|,-1) & j=N+1,\ldots,\cN\coloneqq N+M,
\end{cases}
\end{equation}
and
\begin{equation}\label{eq:Pj}
\mP_j\coloneqq |e_j\rangle\langle f_j| \quad (j=1,\ldots,\cN),
\end{equation}
which allows the two sCM systems \eqref{eq:sCM1a}--\eqref{eq:ejfj} and \eqref{eq:sCM2a}--\eqref{eq:gjhj} to be written as one,
\begin{equation}\label{eq:sCMSHa}
\ddot{a}_j= -2\sum_{k\neq j}^{\cN} (1+r_jr_k)\tr(\mP_j\mP_k)V'(a_j-a_k) \quad (j=1,\ldots,\cN)
\end{equation}
and
\begin{equation}\label{eq:sCMSHb}
\begin{split}
|\dot{e}_j\rangle =&\; \ii\sum_{k\neq j}^{\cN}(1+r_jr_k) \mP_k |e_j\rangle V(a_j-a_k), \\ 
\langle \dot{f}_j|=&\; -\ii \sum_{k\neq j}^{\cN}(1+r_jr_k) \langle f_j|\mP_k    V(a_j-a_k).
\end{split} \quad (j=1,\ldots,\cN)
\end{equation}
and 
\begin{equation}\label{eq:fjejgen}
\langle f_j|e_j\rangle=1 \quad (j=1,\ldots,\cN).
\end{equation}
We note that by \eqref{eq:Pj} and \eqref{eq:fjejgen}, each $\mP_j$ is rank-one projector, i.e., $\mP_j^2=\mP_j$ and $\mathrm{tr}\,\mP_j=1$.

Moreover, because each $|e_j\rangle$ and $\langle f_j|$ for $j=1,\ldots,\cN$ is nonzero by \eqref{eq:fjejgen}, \eqref{eq:BT} is equivalent to the system obtained by left-multiplying the first set of equations in \eqref{eq:BT} by $|e_j\rangle$ and right-multiplying the second set of equations in \eqref{eq:BT} by $\langle f_j|$,
\begin{equation}\label{eq:BTSH}
\dot{a}_j\mP_j=\ii\sum_{k\neq j}^{\cN} r_k \big((1+r_j)\mP_j\mP_k+(1-r_j)\mP_k\mP_j\big)\alpha(a_j-a_k) \quad (j=1,\ldots,\cN),
\end{equation}
where we have used \eqref{eq:SH}--\eqref{eq:Pj}. By taking the trace of \eqref{eq:BTSH}, using \eqref{eq:fjejgen}, we obtain
\begin{equation}\label{eq:BTSH2}
\dot{a}_j=\ii\sum_{k\neq j}^{\cN} r_k \big((1+r_j)\tr(\mP_j\mP_k)+(1-r_j)\tr(\mP_j\mP_k)\big)\alpha(a_j-a_k) \quad (j=1,\ldots,\cN),
\end{equation}
which is equivalent to \eqref{eq:BTscalar} via the notation \eqref{eq:SH}--\eqref{eq:Pj} and the identity $\tr(\mP_j\mP_k)=\langle f_j|e_k\rangle \langle f_k|e_j\rangle$.

\subsection{Proof}\label{sec:mainproof}

Let $\{a_j,|e_j\rangle,\langle f_j|\}_{j=1}^{\cN}$ be the maximal solution of \eqref{eq:sCMSHb} and \eqref{eq:BTSH2} (equivalent to \eqref{eq:sCM1b}, \eqref{eq:sCM2b}, and \eqref{eq:BTscalar}) with the given initial data on a maximal interval $[0,\tau)$. 
It follows from \eqref{eq:BTSH2} that
\begin{equation}\label{eq:BTSH3}
\dot{a}_j\mP_j=\ii\sum_{k\neq j}^{\cN} r_k \big((1+r_j)\tr(\mP_j\mP_k)\mP_j+(1-r_j)\tr(\mP_j\mP_k)\mP_j\big)\alpha(a_j-a_k) \quad (j=1,\ldots,\cN),
\end{equation}
holds on this same interval. The BT \eqref{eq:BTSH} will hold on $[0,\tau)$ provided that the difference of the right-hand sides of \eqref{eq:BTSH3} and \eqref{eq:BTSH} vanishes for each $j=1,\ldots,\cN$; this difference is given up to an overall multiplicative constant by
\begin{equation}\label{eq:Rj}
\mR_j\coloneqq \sum_{k\neq j}^{\cN} r_k \big((1+r_j)\mP_j(\tr(\mP_j\mP_k)-\mP_k)+(1-r_j)(\tr(\mP_j\mP_k)-\mP_k)\mP_j\big)\alpha(a_j-a_k) \quad (j=1,\ldots,\cN). 
\end{equation}
Because \eqref{eq:BTSH} is assumed to hold at $t=0$, we have $\mR_j(0)=0$ for $j=1,\ldots,\cN$. We now consider the time evolution of the $\mR_j$. 

By differentiating \eqref{eq:Rj} with respect to time, we find
\begin{equation}\label{eq:Rjdot}
\dot{\mR}_j=\mC_1+\mC_2+\mC_3+\mC_4+\mC_5, 
\end{equation}
where
\begin{equation}\label{eq:Cj}
\begin{split}
 \mC_1\coloneqq &\; \sum_{k\neq j}^{\cN} r_k \big((1+r_j)\dot{\mP}_j(\tr(\mP_j\mP_k)-\mP_k)+(1-r_j)(\tr(\mP_j\mP_k)-\mP_k)\dot{\mP}_j\big)\alpha(a_j-a_k), \\
 \mC_2 \coloneqq &\; \sum_{k\neq j}^{\cN} r_k\bigg((1+r_j)\mP_j \frac{\mathrm{d}}{\mathrm{d}t}\bigg(\tr(\mP_j\mP_k)\bigg)+(1-r_j)\frac{\mathrm{d}}{\mathrm{d}t}\bigg(\tr(\mP_j\mP_k)\bigg)\mP_j\bigg)\alpha(a_j-a_k), \\
\mC_3 \coloneqq &\; -\sum_{k\neq j}^{\cN} r_k \big((1+r_j)\mP_j\dot{\mP}_k+(1-r_j)\dot{\mP}_k\mP_j\big)\alpha(a_j-a_k),   \\
 \mC_4\coloneqq &\; -\sum_{k\neq j}^{\cN} r_k \big((1+r_j)\mP_j(\tr(\mP_j\mP_k)-\mP_k)+(1-r_j)(\tr(\mP_j\mP_k)-\mP_k)\mP_j\big)\dot{a}_jV(a_j-a_k), \\
\mC_5\coloneqq &\; \sum_{k\neq j}^{\cN} r_k \big((1+r_j)\mP_j(\tr(\mP_j\mP_k)-\mP_k)+(1-r_j)(\tr(\mP_j\mP_k)-\mP_k)\mP_j\big)\dot{a}_kV(a_j-a_k),
\end{split}
\end{equation}
where we have used $\alpha'(z)=-V(z)$ \eqref{eq:alphatoV}.

We compute each quantity $\mC_1,\ldots,\mC_5$ in turn. In doing so, we use the following time evolution equation for $\mP_j$, 
\begin{equation}\label{eq:Pjdot}
\dot{\mP}_j= -\ii \sum_{k\neq j}^{\cN} (1+r_jr_k) [\mP_j,\mP_k]V(a_j-a_k) \quad (j=1,\ldots,\cN),
\end{equation}
where $[\cdot,\cdot]$ is the commutator, which follows from \eqref{eq:SH} and \eqref{eq:sCMSHb}. We will also use the identity
\begin{equation}\label{eq:trId}
\mP_j\mA \mP_j=\tr(\mP_j\mA)\mP_j,
\end{equation}
valid for arbitrary $\mA\in \cV\otimes \cV^*$, at two points below. 

To compute $\mC_1$ in \eqref{eq:Cj}, we first insert \eqref{eq:Pjdot} to find
\begin{align}\label{eq:A1}
\mC_1=&\; -\ii\sum_{k\neq j}^{\cN}\sum_{l\neq j}^{\cN}\bigg( r_k(1+r_jr_l) \big( (1+r_j)[\mP_j,\mP_l](\tr(\mP_j\mP_k)-\mP_k)+(1-r_j)(\tr(\mP_j\mP_k)-\mP_k)[\mP_j,\mP_l]\bigg) \nonumber \\
&\; \phantom{-2\ii\sum_{k\neq j}^{\cN}\sum_{l\neq k}^{\cN}\bigg(} \times \alpha(a_j-a_k)V(a_j-a_l).
\end{align}
To proceed, we write
\begin{equation}\label{eq:AId1}
\begin{split}
[\mP_j,\mP_l](\tr(\mP_j\mP_k)-\mP_k)=&\;[\mP_j(\tr(\mP_j\mP_k)-\mP_k),\mP_l]+\mP_j[\mP_k,\mP_l], \\
(\tr(\mP_j\mP_k)-\mP_k)[\mP_j,\mP_l]=&\;[(\tr(\mP_j\mP_k)-\mP_k)\mP_j,\mP_l]+[\mP_k,\mP_l]\mP_j
\end{split}
\end{equation}
and, using \eqref{eq:Rj}, $(1\pm r_j)^2=2(1\pm r_j)$, and $(1+r_j)(1-r_j)=0$,
\begin{equation}\label{eq:AId2}
\begin{split}
(1+r_j)\mR_j=&\;  2\sum_{k\neq j}^{\cN} r_k(1+r_j)\mP_j(\tr(\mP_j\mP_k)-\mP_k)\alpha(a_j-a_k), \\
(1-r_j)\mR_j=&\;  2\sum_{k\neq j}^{\cN} r_k(1-r_j)(\tr(\mP_j\mP_k)-\mP_k)\mP_j\alpha(a_j-a_k).
\end{split}
\end{equation}
Using \eqref{eq:AId1}--\eqref{eq:AId2} in \eqref{eq:A1} gives
\begin{align}\label{eq:A2}
\mC_1=&\; \ii \sum_{l\neq j}^{\cN} (1+r_jr_l)[\mP_l,\mR_j]V(a_j-a_l) \nonumber \\
&\; -\ii \sum_{k\neq j}^{\cN}\sum_{l\neq j,k}^{\cN} r_k(1+r_jr_l)\big((1+r_j)\mP_j[\mP_k,\mP_l]+(1-r_j)[\mP_k,\mP_l]\mP_j\big)\alpha(a_j-a_k)V(a_j-a_l).
\end{align}

To compute $\mC_2$ in \eqref{eq:Cj}, we start with, using \eqref{eq:Pjdot},
\begin{align}\label{eq:Trdot1}
\frac{\mathrm{d}}{\mathrm{d}t}\tr(\mP_j\mP_k)=&\; \tr(\dot{\mP}_j\mP_k)+\tr(\mP_j\dot{\mP}_k) \nonumber \\
=&\; -\ii \sum_{l\neq j}^{\cN}(1+r_jr_l) \tr([\mP_j,\mP_l]\mP_k)V(a_j-a_l) -\ii\sum_{l\neq k}^{\cN}(1+r_kr_l) \tr(\mP_j[\mP_k,\mP_l])V(a_k-a_l)  
\end{align}
Using the identity $\tr([\mP_j,\mP_l]\mP_k)= -\tr(\mP_j[\mP_k,\mP_l])$ and noting that the $l=k$ term of the first sum and $l=j$ term of the second sum in \eqref{eq:Trdot1} cancel, we find
\begin{equation}\label{eq:Trdot2}
\frac{\mathrm{d}}{\mathrm{d}t}\tr(\mP_j\mP_k)=  \ii\sum_{l\neq j,k}^{\cN} \tr(\mP_j[\mP_k,\mP_l]) \big((1+r_jr_l)V(a_j-a_l)-(1+r_kr_l)V(a_k-a_l)\big).
\end{equation}
Inserting \eqref{eq:Trdot2} into $\mC_2$ in \eqref{eq:Cj} yields
\begin{align}\label{eq:B}
\mC_2=&\; \ii\sum_{k\neq j}^{\cN}\sum_{l\neq j,k}^{\cN}r_k(1+r_jr_l)\tr(\mP_j[\mP_k,\mP_l])\mP_j\big( (1+r_j)+(1-r_j)\big)\alpha(a_j-a_k)V(a_j-a_l) \nonumber  \\
&\; - \ii\sum_{k\neq j}^{\cN}\sum_{l\neq j,k}^{\cN}(r_k+r_l)\tr(\mP_j[\mP_k,\mP_l])\mP_j\big( (1+r_j)+(1-r_j)\big)\alpha(a_j-a_k)V(a_k-a_l).
\end{align}

Next, inserting \eqref{eq:Pjdot} into $\mC_3$ in \eqref{eq:Cj} gives
\begin{align}\label{eq:C}
\mC_3=&\; \ii\sum_{k\neq j}^{\cN}\sum_{l\neq k}^{\cN} r_k(1+r_kr_l)\big( (1+r_j)\mP_j[\mP_k,\mP_l]+(1-r_j)[\mP_k,\mP_l]\mP_j)\alpha(a_j-a_k)V(a_k-a_l) \nonumber \\
=&\; \ii \sum_{k\neq j}^{\cN} (r_j+r_k)\big((1+r_j)\mP_j[\mP_k,\mP_j]+(1-r_j)[\mP_k,\mP_j]\mP_j)\alpha(a_j-a_k)V(a_j-a_k) \nonumber \\
&\; + \ii\sum_{k\neq j}^{\cN}\sum_{l\neq j,k}^{\cN} (r_k+r_l)\big( (1+r_j)\mP_j[\mP_k,\mP_l]+(1-r_j)[\mP_k,\mP_l]\mP_j)\alpha(a_j-a_k)V(a_k-a_l).
\end{align}

It follows from \eqref{eq:A2}, \eqref{eq:B} with \eqref{eq:trId}, and \eqref{eq:C} that
\begin{align}\label{eq:ABC1}
&\mC_1+\mC_2+\mC_3= \nonumber \\
&\ii\sum_{k\neq j}^{\cN} (1+r_jr_k)[\mP_k,\mR_j]V(a_j-a_k) \nonumber \\
&+ \ii \sum_{k\neq j}^{\cN} (r_j+r_k)\big((1+r_j)\mP_j[\mP_k,\mP_j]+(1-r_j)[\mP_k,\mP_j]\mP_j)\alpha(a_j-a_k)V(a_j-a_k) \nonumber \\
& +\ii\sum_{k\neq j}^{\cN}\sum_{l\neq j,k}^{\cN}r_k(1+r_jr_l)\big( (1+r_j)\mP_j[\mP_k,\mP_l](\mP_j-1)+(1-r_j)(\mP_j-1)[\mP_k,\mP_l]\mP_j\big)\alpha(a_j-a_k)V(a_j-a_l) \nonumber \\
&- \ii\sum_{k\neq j}^{\cN}\sum_{l\neq j,k}^{\cN} (r_k+r_l)\big((1+r_j)\mP_j[\mP_k,\mP_l](\mP_j-1)+(1-r_j)(\mP_j-1)[\mP_k,\mP_l]\mP_j\big) \alpha(a_j-a_k)V(a_k-a_l).
\end{align}

For future convenience, we will rewrite the final two lines of \eqref{eq:ABC1}. By swapping indices $k\leftrightarrow l$ and using $(1+r_jr_k)(1\pm r_j)=(1\pm r_j)(1\pm r_k)$, we write the penultimate line of \eqref{eq:ABC1} as
\begin{align}\label{eq:rewrite1}
&\ii\sum_{k\neq j}^{\cN}\sum_{l\neq j,k}^{\cN}r_k(1+r_jr_l)\big( (1+r_j)\mP_j[\mP_k,\mP_l](\mP_j-1)+(1-r_j)(\mP_j-1)[\mP_k,\mP_l]\mP_j\big)\alpha(a_j-a_k)V(a_j-a_l) \nonumber \\
&=-\ii\sum_{k\neq j}^{\cN}\sum_{l\neq j,k}^{\cN}r_l\big( (1+r_j)(1+r_k)\mP_j[\mP_k,\mP_l](\mP_j-1)+(1-r_j)(1-r_k)(\mP_j-1)[\mP_k,\mP_l]\mP_j\big)  \nonumber  \\
&\; \phantom{=-\ii\sum_{k\neq j}^{\cN}\sum_{l\neq j,k}^{\cN}}\times \alpha(a_j-a_l)V(a_j-a_k) \nonumber \\
&= -\ii\sum_{k\neq j}^{\cN}\sum_{l\neq j,k}^{\cN} r_l \big((1+r_j)\mP_j[\mP_k,\mP_l](\mP_j-1)+(1-r_j)[\mP_k,\mP_l]\mP_j\big)   \alpha(a_j-a_l)V(a_j-a_k) \nonumber \\
&\phantom{=}  -\ii\sum_{k\neq j}^{\cN}\sum_{l\neq j,k}^{\cN} r_kr_l\big((1+r_j)\mP_j[\mP_k,\mP_l](\mP_j-1)-(1-r_j)[\mP_k,\mP_l]\mP_j\big)   \alpha(a_j-a_l)V(a_j-a_k).
\end{align}
The final line of \eqref{eq:rewrite1} can be written as
\begin{align}\label{eq:rewrite1,2}
&-\ii\sum_{k\neq j}^{\cN}\sum_{l\neq j,k}^{\cN} r_kr_l\big((1+r_j)\mP_j[\mP_k,\mP_l](\mP_j-1)-(1-r_j)[\mP_k,\mP_l]\mP_j\big)   \alpha(a_j-a_l)V(a_j-a_k) \nonumber \\
&= \ii\sum_{k\neq j}^{\cN}\sum_{l\neq j,k}^{\cN} r_kr_l\big((1+r_j)\mP_j\mP_l\mP_k(\mP_j-1)+(1-r_j)\mP_k\mP_l\mP_j\big)  \nonumber \\
&\phantom{= -\ii\sum_{k\neq j}^{\cN}\sum_{l\neq j,k}^{\cN}}\times \big(\alpha(a_j-a_l)V(a_j-a_k)-\alpha(a_j-a_k)V(a_j-a_l)\big),
\end{align}
as can be seen by symmetrizing both sides of the equality. Similarly, the final line of \eqref{eq:ABC1} can be written as
\begin{align}\label{eq:rewrite2}
&- \ii\sum_{k\neq j}^{\cN}\sum_{l\neq j,k}^{\cN} (r_k+r_l)\big((1+r_j)\mP_j[\mP_k,\mP_l](\mP_j-1)+(1-r_j)(\mP_j-1)[\mP_k,\mP_l]\mP_j\big) \alpha(a_j-a_k)V(a_k-a_l) \nonumber \\
&=- \ii\sum_{k\neq j}^{\cN}\sum_{l\neq j,k}^{\cN} r_l \big((1+r_j)\mP_j[\mP_k,\mP_l](\mP_j-1)+(1-r_j)(\mP_j-1)[\mP_k,\mP_l]\mP_j\big) \nonumber \\
&\phantom{=- \ii\sum_{k\neq j}^{\cN}\sum_{l\neq j,k}^{\cN}}\times  \big(\alpha(a_j-a_k)-\alpha(a_j-a_l)\big)V(a_k-a_l),
\end{align}
using also the fact that $V(z)$ is an even function \eqref{eq:parity}.

Hence, after inserting \eqref{eq:rewrite1} with \eqref{eq:rewrite1,2} and \eqref{eq:rewrite2}, \eqref{eq:ABC1} may be written as 
\begin{align}\label{eq:ABC2}
&\mC_1+\mC_2+\mC_3= \nonumber \\
&\ii\sum_{k\neq j}^{\cN} (1+r_jr_k)[\mP_k,\mR_j]V(a_j-a_k) \nonumber \\
&+ \ii \sum_{k\neq j}^{\cN} (r_j+r_k)\big((1+r_j)\mP_j[\mP_k,\mP_j]+(1-r_j)[\mP_k,\mP_j]\mP_j)\alpha(a_j-a_k)V(a_j-a_k) \nonumber \\
& -\ii\sum_{k\neq j}^{\cN}\sum_{l\neq j,k}^{	 \cN}  r_l\big( (1+r_j)\mP_j[\mP_k,\mP_l](\mP_j-1)+(1-r_j)(\mP_j-1)[\mP_k,\mP_l]\mP_j\big) \nonumber \\
& \phantom{ -\ii\sum_{k\neq j}^{\cN}\sum_{l\neq j,k}^{\cN}}  \times \big(\alpha(a_j-a_l)V(a_j-a_k)+\big(\alpha(a_j-a_k)-\alpha(a_j-a_l)\big)V(a_k-a_l)\big)   \nonumber  \\
&+ \ii\sum_{k\neq j}^{\cN}\sum_{l\neq j,k}^{\cN} r_k r_l\big((1+r_j)\mP_j\mP_l\mP_k(\mP_j-1)+(1-r_j)(\mP_j-1)\mP_k\mP_l\mP_j\big) \nonumber \\
&\phantom{+ \ii\sum_{k\neq j}^{\cN}\sum_{l\neq j,k}^{\cN}} \times \big(\alpha(a_j-a_l)V(a_j-a_k)-\alpha(a_j-a_k)V(a_j-a_l)\big).
\end{align}

A key tool in computing $\mC_4$ and $\mC_5$ in \eqref{eq:Cj} is the following relation, which follows from \eqref{eq:BTSH3} and \eqref{eq:Rj},
\begin{equation}\label{eq:ajdotPj2}
\dot{a}_j\mP_j= \ii \mR_j+\ii \sum_{k\neq j}^{\cN} r_k \big((1+r_j)\mP_j\mP_k+(1-r_j)\mP_k\mP_j\big)   \alpha(a_j-a_k).
\end{equation}

First, we insert \eqref{eq:ajdotPj2} into $\mC_4$ in \eqref{eq:Cj},
\begin{align}\label{eq:Dterm1,1}
\mC_4=&\; -\sum_{k\neq j}^{\cN} r_k \big((1+r_j)(\dot{a}_j\mP_j)(\tr(\mP_j\mP_k)-\mP_k)+(1-r_j)(\tr(\mP_j\mP_k)-\mP_k)(\dot{a}_j\mP_j)\big)V(a_j-a_k) \nonumber  \\
=&\;  -\ii \sum_{k\neq j}^{\cN} r_k\big((1+r_j)\mR_j(\tr(\mP_j\mP_k)-\mP_k)+(1-r_j)(\tr(\mP_j\mP_k)-\mP_k)\mR_j\big)V(a_j-a_k) \nonumber \\
&\; -2\ii \sum_{k\neq j}^{\cN} \sum_{l\neq j}^{\cN} r_k r_l\big((1+r_j)\mP_j\mP_l(\tr(\mP_j\mP_k)-\mP_k)+(1-r_j)(\tr(\mP_j\mP_k)-\mP_k)\mP_l\mP_j\big) \nonumber  \\
&\; \phantom{-2\ii \sum_{k\neq j}^{\cN} \sum_{l\neq j}^{\cN}} \times \alpha(a_j-a_l)V(a_j-a_k)
\end{align}
We use \eqref{eq:AId2} to replace the terms in $\mR_j\mP_k$ and $\mP_k\mR_j$ in \eqref{eq:Dterm1,1}, and this leads to
\begin{align}\label{eq:Dterm1,2}
\mC_4 =&\;   -\ii \sum_{k\neq j}^{\cN} r_k\big((1+r_j)\mR_j\tr(\mP_j\mP_k)+(1-r_j)\tr(\mP_j\mP_k)\mR_j\big)V(a_j-a_k) \nonumber  \\
&\;  -2\ii \sum_{k\neq j}^{\cN} \sum_{l\neq j}^{\cN} r_kr_l\big((1+r_j)\mP_j\mP_l\tr(\mP_j\mP_k)+(1-r_j)\tr(\mP_j\mP_k)\mP_l\mP_j\big)\alpha(a_j-a_l)V(a_j-a_k) \nonumber  \\
&\; +2\ii \sum_{k\neq j}^{\cN}\sum_{l\neq j}^{\cN} r_kr_l\big((1+r_j)\mP_j\mP_k\tr(\mP_j\mP_l)+(1-r_j)\tr(\mP_j\mP_l)\mP_k\mP_j\big)\alpha(a_j-a_l)V(a_j-a_k) \nonumber  \\
=&\;   -2\ii \sum_{k\neq j}^{\cN} r_k \tr(\mP_j\mP_k)\mR_jV(a_j-a_k) \nonumber  \\
&\; -2\ii \sum_{k\neq j}^{\cN} \sum_{l\neq j}^{\cN} r_k r_l \big((1+r_j) [\mP_j\mP_k,\mP_j\mP_l]-(1-r_j)[\mP_k\mP_j,\mP_l\mP_j] \big)\alpha(a_j-a_l)V(a_j-a_k),
\end{align}
where we have used \eqref{eq:trId} in the last step.

Next, we insert \eqref{eq:ajdotPj2} into $\mC_5$ in \eqref{eq:Cj},
\begin{align} \label{eq:Dterm2,1}
\mC_5= &\; \sum_{k\neq j}^{\cN}r_k\big((1+r_j)\mP_j(\dot{a}_k\mP_k)(\mP_j-1)+(1-r_j)(\mP_j-1)(\dot{a}_k\mP_k)\mP_j\big)V(a_j-a_k) \nonumber \\
=&\;   \ii  \sum_{k\neq j}^{\cN} r_k\big((1+r_j)\mP_j\mR_k (\mP_j-1) + (1-r_j)(\mP_j-1)\mP_k\mR_j\big)V(a_j-a_k) \nonumber \\
&\; +\ii\sum_{k\neq j}^{\cN} \sum_{l\neq k}^{\cN} (1+r_j)r_l\big((1+r_k)\mP_j\mP_k\mP_l(\mP_j-1)-(1-r_k)\mP_j\mP_l\mP_k(\mP_j-1)\big)\alpha(a_k-a_l)V(a_j-a_k)  \nonumber \\
&\; +\ii \sum_{k\neq j}^{\cN} \sum_{l\neq k}^{\cN} (1-r_j)r_l\big((1+r_k)(\mP_j-1)\mP_k\mP_l\mP_j-(1-r_k)(\mP_j-1)\mP_l\mP_k\mP_j\big)\alpha(a_k-a_l)V(a_j-a_k),
\end{align}
where we have used $r_k(1\pm r_k)=\pm(1\pm r_k)$. The $l=j$ terms of the double sums in \eqref{eq:Dterm2,1} amount to
\begin{align}\label{eq:Dterm2,2}
&\ii \sum_{k\neq j}^{\cN} \big( (1+r_j)(1-r_k)\mP_j\mP_k(\mP_j-1)+(1-r_j)(1+r_k)(\mP_j-1)\mP_k\mP_j\big)      \alpha(a_j-a_k)V(a_j-a_k) \nonumber \\
&=  -\ii \sum_{k\neq j}^{\cN} (r_j+r_k)\big((1+r_j)\mP_j[\mP_k,\mP_j]+(1-r_j)[\mP_k,\mP_j]\mP_j)\alpha(a_j-a_k)V(a_j-a_k) \nonumber \\
&\phantom{=} +2\ii\sum_{k\neq j}^{\cN} \big((1+r_j)\mP_j[\mP_k,\mP_j]+(1-r_j)[\mP_j,\mP_k]\mP_j\big)\alpha(a_j-a_k)V(a_j-a_k) \nonumber \\
&=  -\ii \sum_{k\neq j}^{\cN} (r_j+r_k)\big((1+r_j)\mP_j[\mP_k,\mP_j]+(1-r_j)[\mP_k,\mP_j]\mP_j)\alpha(a_j-a_k)V(a_j-a_k) \nonumber \\ 
&\phantom{=} +2\ii\sum_{k\neq j}^{\cN} r_k \mR_jV(a_j-a_k) \nonumber \\
&\phantom{=} -2\ii\sum_{k\neq j}^{\cN} \sum_{l\neq j,k}^{\cN} r_kr_l\big((1+r_j)\mP_j[\mP_l,\mP_j]+(1-r_j)[\mP_j,\mP_l]\mP_j\big)\alpha(a_j-a_l)V(a_j-a_k), 
\end{align}
where we have used \eqref{eq:Rj} in the second step. Thus, combining \eqref{eq:Dterm2,1} and \eqref{eq:Dterm2,2} and simplifying gives
\begin{align} \label{eq:Dterm2,3}
\mC_5=&\;  2\ii \sum_{k\neq j}^{\cN} r_k \mR_j V(a_j-a_k) +\ii \sum_{k\neq j}^{\cN} r_k \big((1+r_j) r_k\mP_j\mR_k (\mP_j-1) +(1-r_j)(\mP_j-1)\mP_k\mR_j\big)V(a_j-a_k) \nonumber \\
&\; -\ii \sum_{k\neq j}^{\cN} (r_j+r_k)\big((1+r_j)\mP_j[\mP_l,\mP_j]+(1-r_j)[\mP_k,\mP_j]\mP_j)\alpha(a_j-a_k)V(a_j-a_k) \nonumber \\
&\; -2\ii\sum_{k\neq j}^{\cN} \sum_{l\neq j,k}^{\cN} r_kr_l\big((1+r_j)\mP_j[\mP_l,\mP_j]+(1-r_j)[\mP_j,\mP_l]\mP_j\big)\alpha(a_j-a_l)V(a_j-a_k) \nonumber \\
&\; +\ii \sum_{k\neq j}^{\cN} \sum_{l\neq j,k}^{\cN} r_l \big((1+r_j)\mP_j[\mP_k,\mP_l](\mP_j-1)+(1-r_j)(\mP_j-1)[\mP_k,\mP_l]\mP_j\big)\alpha(a_k-a_l)V(a_j-a_k)  \nonumber \\
&\; +\ii \sum_{k\neq j}^{\cN} \sum_{l\neq j,k}^{\cN} r_kr_l \big((1+r_j)\mP_j\{\mP_k,\mP_l\}(\mP_j-1)+(1-r_j)(\mP_j-1)\{\mP_k,\mP_l\}\mP_j\big)\alpha(a_k-a_l)V(a_j-a_k),
\end{align}
where $\{\cdot,\cdot\}$ is the anti-commutator.

By adding \eqref{eq:Dterm1,2} and \eqref{eq:Dterm2,3}, we obtain
\begin{align}\label{eq:D}
&\mC_4+\mC_5= \nonumber \\
& \ii \sum_{k\neq j}^{\cN} r_k\big(2(1-\tr(\mP_j\mP_k))\mR_j+(1+r_j) \mP_j\mR_k (\mP_j-1) + (1-r_j)(\mP_j-1)\mP_k\mR_j \big)V(a_j-a_k) \nonumber \\
& -2\ii \sum_{k\neq j}^{\cN} \sum_{l\neq j,k}^{\cN} r_k r_l \big(1+r_j)[\mP_j\mP_k,\mP_j\mP_l]-(1-r_j)[\mP_k\mP_j,\mP_l\mP_j]    \big)\alpha(a_j-a_l)V(a_j-a_k) \nonumber \\
& -2\ii\sum_{k\neq j}^{\cN}\sum_{l\neq j,k}^{\cN} r_kr_l\big((1+r_j)\mP_j[\mP_l,\mP_j]+(1-r_j)[\mP_j,\mP_l]\mP_j\big)\alpha(a_j-a_l)V(a_j-a_k) \nonumber \\
& +\ii \sum_{k\neq j}^{\cN} \sum_{l\neq j,k}^{\cN} r_l \big((1+r_j)\mP_j[\mP_k,\mP_l](\mP_j-1)+(1-r_j)(\mP_j-1)[\mP_k,\mP_l]\mP_j\big)\alpha(a_k-a_l)V(a_j-a_k) \nonumber \\
&+\ii \sum_{k\neq j}^{\cN} \sum_{l\neq j,k}^{\cN} r_kr_l \big((1+r_j)\mP_j \mP_l \mP_k(\mP_j-1)+(1-r_j)(\mP_j-1)\mP_l\mP_k\mP_j\big) \nonumber \\
&\phantom{+\ii \sum_{k\neq j}^{\cN} \sum_{l\neq j,k}^{\cN}} \times \alpha(a_k-a_l)\big(V(a_j-a_k)-V(a_j-a_l)\big),
\end{align}
where, similarly as in \eqref{eq:rewrite1,2}-\eqref{eq:rewrite2}, we have used symmetry to rewrite the final line. 

By inserting \eqref{eq:ABC2}, \eqref{eq:D}, and the identities
\begin{equation}\label{eq:alphaV}
\alpha(a_j-a_l)V(a_j-a_k)+\big(\alpha(a_j-a_k)-\alpha(a_j-a_l)\big)V(a_k-a_l)-\alpha(a_k-a_l)V(a_j-a_k)=0
\end{equation}
and
\begin{multline}
\alpha(a_j-a_l)V(a_j-a_k)-\alpha(a_j-a_k)V(a_j-a_l)+\alpha(a_k-a_l)\big(V(a_j-a_k)-V(a_j-a_l)\big),   \\ 
=2\alpha(a_j-a_l)V(a_j-a_k),
\end{multline}
each of which can be found by differentiating \eqref{eq:alphaId} with respect to a particular variable and then renaming variables, into \eqref{eq:Rjdot}, we obtain 
\begin{align}\label{eq:ABCD1}
\dot{\mR}_j= &\; \ii\sum_{k\neq j}^{\cN} (1+r_jr_k)[\mP_k,\mR_j]V(a_j-a_k) \nonumber \\
&\; + \ii \sum_{k\neq j}^{\cN} r_k\big(2(1-\tr(\mP_j\mP_k))\mR_j+(1+r_j) \mP_j\mR_k (\mP_j-1) + (1-r_j)(\mP_j-1)\mP_k\mR_j \big)V(a_j-a_k) \nonumber \\
&\; -2\ii \sum_{k\neq j}^{\cN} \sum_{l\neq j,k}^{\cN} r_k r_l \big((1+r_j)[\mP_j\mP_k,\mP_j\mP_l]-(1-r_j)[\mP_k\mP_j,\mP_l\mP_j]\big)\alpha(a_j-a_l)V(a_j-a_k) \nonumber \\
&\; -2\ii\sum_{k\neq j}^{\cN}\sum_{l\neq j,k}^{\cN} r_kr_l\big((1+r_j)\mP_j[\mP_l,\mP_j]+(1-r_j)[\mP_j,\mP_l]\mP_j\big)\alpha(a_j-a_l)V(a_j-a_k) \nonumber \\
&\; +2\ii \sum_{k\neq j}^{\cN} \sum_{l\neq j,k}^{\cN} r_kr_l \big((1+r_j)\mP_j \mP_l \mP_k(\mP_j-1)+(1-r_j)(\mP_j-1)\mP_k\mP_l\mP_j\big) \alpha(a_j-a_l)V(a_j-a_k).
\end{align}
The final two lines of \eqref{eq:ABCD1} can be combined into 
\begin{align}\label{eq:ABCD1,2}
&2\ii \sum_{k\neq j}^{\cN} \sum_{l\neq j,k}^{\cN} r_kr_l \big((1+r_j)\mP_j \mP_l (\mP_k-1)(\mP_j-1)+(1-r_j)(\mP_j-1)(\mP_k-1)\mP_l\mP_j\big) \alpha(a_j-a_l)V(a_j-a_k) \nonumber \\
&=-2\ii \sum_{k\neq j}^{\cN} \sum_{l\neq j,k}^{\cN} r_kr_l \big((1+r_j)[(\mP_k-1)(\mP_j-1),\mP_j \mP_l ]-(1-r_j)[(\mP_j-1)(\mP_k-1),\mP_l\mP_j]\big) \nonumber \\
&\phantom{=2\ii \sum_{k\neq j}^{\cN}\sum_{l\neq j,k}^{\cN}}  \times\alpha(a_j-a_l)V(a_j-a_k),
\end{align}
where we have used $(\mP_j-1)\mP_j=0=\mP_j(\mP_j-1)$. By further combining \eqref{eq:ABCD1,2} with the third line of \eqref{eq:ABCD1}, we find
\begin{align}\label{eq:ABCD2}
\dot{\mR}_j =&\;  \ii\sum_{k\neq j}^{\cN} (1+r_jr_k)[\mP_k,\mR_j]V(a_j-a_k) \nonumber \\
&\; + \ii \sum_{k\neq j}^{\cN} r_k\big(2(1-\tr(\mP_j\mP_k))\mR_j+(1+r_j) \mP_j\mR_k (\mP_j-1) + (1-r_j)(\mP_j-1)\mP_k\mR_j \big)V(a_j-a_k) \nonumber \\
&\; -2\ii \sum_{k\neq j}^{\cN}\sum_{l\neq j,k}^{\cN} r_kr_l \big((1+r_j) [\mP_k(\mP_j-1)+\mP_j(\mP_k-1),\mP_j\mP_l] \nonumber \\
& \phantom{\;+2\ii \sum_{k\neq j}^{\cN}\sum_{l\neq j,k}^{\cN}r_kr_l \big(} -(1-r_j)[\mP_k(\mP_j-1)+\mP_j(\mP_k-1),\mP_l\mP_j]\big) \alpha(a_j-a_l)V(a_j-a_k).
\end{align}
The double sum in \eqref{eq:ABCD2} can be written as 
\begin{align}\label{eq:ABCD3}
&-\ii \sum_{k\neq j}^{\cN}r_k \Bigg[   \mP_j(\mP_k-1)+\mP_k(\mP_j-1), 2\sum_{l\neq j,k}^{\cN} r_l(1+r_j)  \mP_j\mP_l\alpha(a_j-a_l)\Bigg] V(a_j-a_k) \nonumber \\
&+ \ii \sum_{k\neq j}^{\cN} r_k\Bigg[  \mP_j(\mP_k-1)+\mP_k(\mP_j-1), 2\sum_{l\neq j,k}^{\cN} r_l (1-r_j) \mP_l\mP_j\alpha(a_j-a_l)\Bigg] V(a_j-a_k).
\end{align}
Using that
\begin{equation}
[\mP_j(\mP_k-1)+\mP_k(\mP_j-1),\mP_j\mP_k]=0=[\mP_j(\mP_k-1)+\mP_k(\mP_j-1),\mP_k\mP_j]
\end{equation}
and
\begin{equation}
[\mP_j(\mP_k-1)+\mP_k(\mP_j-1),\mP_j]=0,
\end{equation}
we write \eqref{eq:ABCD3} as
\begin{align}\label{eq:ABCD4}
&\ii \sum_{k\neq j}^{\cN}r_k \Bigg[   \mP_j(\mP_k-1)+\mP_k(\mP_j-1), 2\sum_{l\neq j}^{\cN} r_l(1+r_j)  \mP_j(\tr(\mP_j\mP_l)-\mP_l)\alpha(a_j-a_l)\Bigg] V(a_j-a_k) \nonumber \\
& - \ii \sum_{k\neq j}^{\cN} r_k\Bigg[  \mP_j(\mP_k-1)+\mP_k(\mP_j-1), 2\sum_{l\neq j}^{\cN} r_l (1-r_j) (\tr(\mP_j\mP_l)-\mP_l)\mP_j\alpha(a_j-a_l)\Bigg] V(a_j-a_k).
\end{align}
We recognize the second arguments of the commutators in \eqref{eq:ABCD4} as the right-hand sides of \eqref{eq:AId2}. Replacing the double sum in \eqref{eq:ABCD2} by \eqref{eq:ABCD4} with \eqref{eq:AId2} gives
\begin{align}\label{eq:Rjdot2}
\dot{\mR}_j=&\;  \ii\sum_{k\neq j}^{\cN} (1+r_jr_k)[\mP_k,\mR_j]V(a_j-a_k) \nonumber \\
&\; + \ii \sum_{k\neq j}^{\cN} r_k\big(2(1-\tr(\mP_j\mP_k))\mR_j+(1+r_j) \mP_j\mR_k (\mP_j-1) + (1-r_j)(\mP_j-1)\mR_k\mP_j \big)V(a_j-a_k) \nonumber \\
&\; +2\ii\sum_{k\neq j}^{\cN} r_j r_k  [\mP_j(\mP_k-1)+\mP_k(\mP_j-1), \mR_j]V(a_j-a_k),
\end{align}
valid for $j=1,\ldots,\cN$. 

We have constructed a system of linear homogenous differential equations obeyed by $\{\mR_j\}_{j=1}^{\cN}$, with coefficients in known variables $\{a_j,\mP_j=|e_j\rangle\langle f_j|\}_{j=1}^{\cN}$. Because $\{a_j,|e_j\rangle,\langle f_j|\}_{j=1}^{\cN}$ is a maximal solution, each solution variable is finite and \eqref{eq:ajak} holds on $[0,\tau)$, so that each coefficient in \eqref{eq:Rjdot2} is finite on $[0,\tau)$. Equipped with the initial conditions $\{\mR_j(0)=0\}_{j=1}^{\cN}$, we conclude that $\{\mR_j(t)=0\}_{j=1}^{\cN}$ is the unique solution of \eqref{eq:Rjdot2} on $[0,\tau)$. It follows that \eqref{eq:BTSH} and consequently \eqref{eq:BT} is satisfied on $[0,\tau)$.

We have shown that $\{a_j,|e_j\rangle,\langle f_j|\}_{j=1}^{\cN}$ solves the system of equations \eqref{eq:sCM1b}, \eqref{eq:sCM2b}, and \eqref{eq:BT} on $[0,\tau)$. We claim that this solution is unique. A necessary condition for the solvability of \eqref{eq:sCM1b}, \eqref{eq:sCM2b}, and \eqref{eq:BT} is the solvability of the reduced system \eqref{eq:sCM1b}, \eqref{eq:sCM2b}, and \eqref{eq:BTscalar}, for which we have constructed a unique solution on $[0,\tau)$. It follows that $\{a_j,|e_j\rangle,\langle f_j|\}_{j=1}^{\cN}$ is the only possible solution of \eqref{eq:sCM1b}, \eqref{eq:sCM2b}, and \eqref{eq:BT} on $[0,\tau)$. This completes the proof.

\appendix

\section{Special functions}\label{app:special}

We recall three basic properties of the special functions $\alpha(z)$ and $V(z)$ defined in \eqref{eq:alpha} and \eqref{eq:V}, respectively.

First, the functions $\alpha(z)$ and $V(z)$ are odd and even functions of $z$, respectively:
\begin{equation}\label{eq:parity}
\alpha(-z)=-\alpha(z), \qquad V(-z)=V(z) \quad (z\in \C)
\end{equation}

Second, the functions $\alpha(z)$ and $V(z)$ are related by differentiation,
\begin{equation}\label{eq:alphatoV}
\alpha'(z)=-V(z) \quad (z\in \C).
\end{equation}

Third, the function $\alpha(z)$ satisfies the identity
\begin{equation}\label{eq:alphaId}
\alpha(a-b)\alpha(a-c)=\alpha(b-c)\big(\alpha(a-b)-\alpha(a-c)\big)+C \quad (a,b,c\in \C),
\end{equation}
where
\begin{equation}
C\coloneqq \begin{cases}
0 & \text{(case I)} \\
(\pi/L)^2 & \text{(case II)} \\
-(\pi/2\delta)^2 & \text{(case III)}.
\end{cases}
\end{equation}

\section*{Acknowledgements}

I would like to thank E. Langmann and J. Lenells for useful discussions and collaboration on closely related projects. This work was supported by the Olle Engkvist Foundation, Grant 211-0122.

\section*{Data Availability}

Data sharing not applicable to this article as no datasets were generated or analysed during the current study.

\bibliography{BT.bib}

\begin{thebibliography}{10}
\providecommand{\natexlab}[1]{#1}
\providecommand{\url}[1]{\texttt{#1}}
\expandafter\ifx\csname urlstyle\endcsname\relax
  \providecommand{\doi}[1]{doi: #1}\else
  \providecommand{\doi}{doi: \begingroup \urlstyle{rm}\Url}\fi

\bibitem[Gibbons and Hermsen(1984)]{gibbons1984}
J.~Gibbons and T.~Hermsen.
\newblock {A generalisation of the Calogero-Moser system}.
\newblock \emph{Physica D}, 11:\penalty0 337--348, 1984.
\newblock URL \url{https://doi.org/10.1016/0167-2789(84)90015-0}.

\bibitem[Wojciechowski(1985)]{wojciechowski1985}
S.~Wojciechowski.
\newblock {An integrable marriage of the Euler equations to the Calogero-Moser
  system}.
\newblock \emph{Physica D}, 28:\penalty0 101--103, 1985.
\newblock ISSN 0167-2789.
\newblock URL \url{https://doi.org/10.1016/0375-9601(85)90432-3}.

\bibitem[Krichever et~al.(1995)Krichever, Babelon, Billey, and
  Talon]{krichever1995}
I.~Krichever, O.~Babelon, E.~Billey, and M.~Talon.
\newblock {Spin generalization of the Calogero-Moser system and the matrix KP
  equation}.
\newblock In S.P Novikov, editor, \emph{Topics in Topology and Mathematical
  Physics}, volume 170, pages 83--120. American Mathematical Society, 1995.

\bibitem[Gibbons et~al.(1983)Gibbons, Hermsen, and Wojciechowski]{gibbons1983}
J.~Gibbons, T.~Hermsen, and S.~Wojciechowski.
\newblock A {B}{\"a}cklund transformation for a generalised {C}alogero-{M}oser
  system.
\newblock \emph{Phys. Lett. A}, 94:\penalty0 251--253, 1983.
\newblock URL \url{https://doi.org/10.1016/0375-9601(83)90710-7}.

\bibitem[Zabrodin(2019)]{zabrodin2019}
A.~Zabrodin.
\newblock {Time discretization of the spin Calogero-Moser model and the
  semi-discrete matrix KP hierarchy}.
\newblock \emph{J. Math. Phys.}, 60:\penalty0 033502, 2019.
\newblock URL \url{https://doi.org/10.1063/1.5081021}.

\bibitem[Prokofev and Zabrodin(2020)]{prokofev2020}
V.V. Prokofev and A.V. Zabrodin.
\newblock {Matrix Kadomtsev-Petviashvili Hierarchy and Spin Generalization of
  Trigonometric Calogero-Moser Hierarchy}.
\newblock \emph{Proc. Steklov Inst.}, 309\penalty0 (1):\penalty0 225--239,
  2020.
\newblock URL \url{https://doi.org/10.1134/S0081543820030177}.

\bibitem[Berntson et~al.(2022)Berntson, Langmann, and
  Lenells]{berntsonlangmann2022}
B.K. Berntson, E.~Langmann, and J.~Lenells.
\newblock {Spin generalizations of the Benjamin-Ono equation}.
\newblock \emph{Lett. Math. Phys.}, 112\penalty0 (50), 2022.
\newblock URL \url{https://doi.org/10.1007/s11005-022-01540-3}.

\bibitem[Berntson et~al.()Berntson, Langmann, and
  Lenells]{berntsonlangmann2022elliptic}
B.K. Berntson, E.~Langmann, and J.~Lenells.
\newblock Elliptic soliton solutions of the spin non-chiral intermediate long
  wave equation.
\newblock In preparation.

\bibitem[Dirac(1939)]{dirac1939}
P.A.M. Dirac.
\newblock A new notation for quantum mechanics.
\newblock \emph{Math. Proc. Cambridge Philos. Soc.}, 35:\penalty0 416--418,
  1939.
\newblock URL \url{https://doi.org/10.1017/S0305004100021162}.

\bibitem[Hartman(1982)]{hartman1982}
P.~Hartman.
\newblock \emph{Ordinary differential equations}.
\newblock Birkh\"{a}user, Boston, Massachusetts, reprint of the second edition,
  1982.

\end{thebibliography}

\end{document}